\documentclass[preprint,journal]{arxiv} 

\ifpdf
  \pdfoutput=1\relax                   
  \usepackage{graphicx}                
  \DeclareGraphicsExtensions{.pdf,.png,.jpg,.jpeg} 
\else
  \usepackage{graphicx}                
  \DeclareGraphicsExtensions{.pdf}     
\fi%

\graphicspath{{figures/}{pictures/}{images/}{./}} 

\usepackage{microtype}                 
\PassOptionsToPackage{warn}{textcomp}  
\usepackage{textcomp}                  
\usepackage{mathptmx}                  
\usepackage{times}                     
\usepackage{cite}                      
\usepackage{tabu}                      
\usepackage{booktabs}                  

\onlineid{1039}

\newif\ifmaindocument

\maindocumenttrue

\ifmaindocument
\vgtccategory{Research}
\vgtcpapertype{technique}

\title{Photon Field Networks for Dynamic\\Real-Time Volumetric Global Illumination}

\author{%
  \authororcid{David~Bauer}{0000-0002-1327-3054},
  \authororcid{Qi~Wu}{0000-0002-1825-0097},
  and~\authororcid{Kwan-Liu~Ma}{0000-0000-0000-0000}
}

\authorfooter{
  \item Authors are with the University of California, Davis.
  E-mail: {\normalfont\{}davbauer\,$|$\,qadwu\,$|$\,klma{\normalfont\}}@ucdavis.edu\,.
}

\abstract{Volume data is commonly found in many scientific disciplines, like medicine, physics, and biology. Experts rely on robust scientific visualization techniques to extract valuable insights from the data. Recent years have shown path tracing to be the preferred approach for volumetric rendering, given its high levels of realism. However, real-time volumetric path tracing often suffers from stochastic noise and long convergence times, limiting interactive exploration. In this paper, we present a novel method to enable real-time global illumination for volume data visualization. We develop \textit{Photon Field Networks}---a phase-function-aware, multi-light neural representation of indirect volumetric global illumination. The fields are trained on multi-phase photon caches that we compute a priori. Training can be done within seconds, after which the fields can be used in various rendering tasks. To showcase their potential, we develop a custom neural path tracer, with which our photon fields achieve interactive framerates even on large datasets. We conduct in-depth evaluations of the method's performance, including visual quality, stochastic noise, inference and rendering speeds, and accuracy regarding illumination and phase function awareness. Results are compared to ray marching, path tracing and photon mapping. Our findings show that \textit{Photon Field Networks} can faithfully represent indirect global illumination across the phase spectrum while exhibiting less stochastic noise and rendering at a significantly faster rate than traditional methods.}


\CCScatlist{
  \CCScatTwelve{Computing methodologies}{Rendering};
  \CCScatTwelve{omputing methodologies}{Neural networks};
  \CCScatTwelve{Computing methodologies}{Volumetric models};
}

\teaser{
  \centering
  \includegraphics[width=\textwidth]{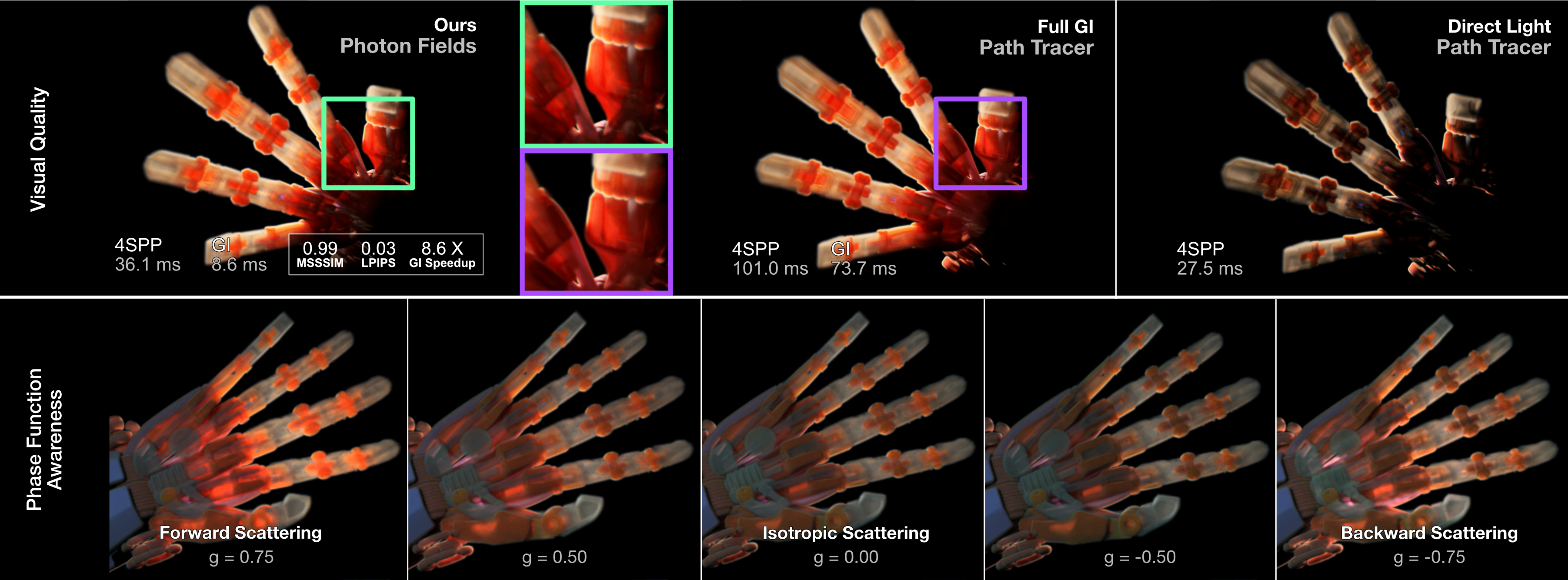}
  \caption{Our photon field networks replace the global illumination term of the rendering equation. With our approach we are able to achieve comparable results for volumetric rendering in a fraction of the time it takes a conventional path tracer. The photon fields are light-weight and can be trained in seconds.}
  \label{fig:teaser}
}

\else
    \input{sections/preamble_supplemental}
\fi

\usepackage{multirow}
\usepackage{algorithm}
\usepackage{algpseudocode}
\usepackage{amsmath}
\usepackage{amssymb}
\usepackage{tabularx}
\usepackage{caption}
\usepackage{subcaption}
\usepackage{graphicx}
\usepackage{soul}

\begin{document}



\maketitle

\ifmaindocument
    
    \section{Introduction}

Volume data is generated in various scientific settings like large-scale physical simulations, medical and biological scans, or astronomical observations. Researchers in these domains rely on robust and high-fidelity rendering techniques to visualize and extract valuable insights from their data. When it comes to visualizing the data, there is a wide array of rendering techniques ranging from unshaded ray marching to full global illumination path tracing. In many applications, simple local illumination rendering is not ideal for visualizing complex volume data, making it hard to study the shape, size, and spatial relations of features in the data. Providing higher levels of realism and visual fidelity generally aids analysis. However, a rendering technique's visual quality and realism closely correlate with its computational cost as it involves casting additional rays to sample the environment. On the other hand, one of the main objectives when developing visualization applications for researchers is providing systems with high interactivity and image quality to facilitate analysis. These requirements put many advanced global illumination effects out of reach for interactive applications. This is especially true for path-traced applications where several light bounces per ray must be computed to achieve realistic global illumination. At the same time, tracing these paths is prone to stochastic noise, requiring numerous samples per pixel (SPP) and smart sampling decisions to achieve acceptable image quality.

In this work, we introduce a method that meets the interactivity requirement of scientific visualization applications without sacrificing realistic global illumination effects. To enable real-time volumetric global illumination for scientific data, we develop the concept of photon fields that allow us to represent phase-dependent, volumetric global illumination as a neural field, through which we can parameterize scene radiance using 6-dimensional coordinates. At the same time, the field can be queried in real time at several SPP. This eliminates the need to construct long paths through the volume and reduces the complexity to a single query to the photon field plus the computation of direct illumination. We implement a neural path tracer that combines traditional direct illumination with photon field queries to showcase our method.

With the implementation, we conduct an extensive evaluation of the photon field renderer involving a priori tracing and training performance, rendering speed, image quality, and sensitivity to phase function changes. The test results show that photon field networks can faithfully represent implicit phase-function-dependent scene radiance while rendering significantly faster and with less stochastic noise than conventional methods.

    \section{Related Work}
Our work is related to topics in volume rendering, global illumination for volume rendering, and deep learning methods in computer graphics. In this section, we discuss related works in these fields.\\

\textbf{Volumetric Global Illumination}
In 1995, Max et al.~\cite{max1995optical} expertly compiled several optical models that allow the realistic rendering of volumetric data. Ever since, these techniques have been at the core of work in the scientific visualization community. Today, two common rendering methods found in scientific rendering are direct volume rendering via ray marching and ray tracing using volumetric path tracing algorithms. Although different in their approach and areas of use, both directions have ways of introducing global illumination (GI) effects such as shadows, ambient occlusion, or, more generally, multiple-scattering effects. In the context of scientific visualization, adding such effects aids users in analyzing their data as these techniques help bring out details and accentuate the spatial properties of the dataset.
On the other hand, these effects can also serve purely aesthetic ends when the purpose of the visualization is to illustrate, promote, or communicate in a general-public setting.

One of the challenges of volumetric global illumination is performance. Rendering methods tend to require large numbers of samples and computations to achieve high-fidelity rendering. Over the years, extensive work has been done to approximate these GI effects while saving computational resources. A notable method to add depth to volumetric rendering is ambient occlusion~\cite{diaz2010real, ropinski2008interactive, hernell2009local, ruiz2008obscurance, schott2009directional, vsolteszova2010multidirectional, kroes2015smooth}, which approximates GI by sampling light extinction in a local neighborhood. The logical next step on the realism scale is to compute global shadows. These consider the global attenuation between samplings points and light sources and represent single-scattering illumination reaching a point. Efficient computation of global shadows requires optimizations over naïve extinction sampling. Notable examples include cone tracing~\cite{shih2016parallel}, plane sweep~\cite{sunden2011image}, half-angle slicing~\cite{kniss2002interactive, kniss2003model}, and light volumes~\cite{zhang2013fast}. Finally, with volumetric path tracing, ray tracing, and photon mapping, we can compute the full multiple-scattering GI model~\cite{kroes2012exposure, paladini2015optimization, liu2016progressive, dappa2016cinematic, zhang2013real}. However, the computational cost of the algorithms increases when high-resolution data and complex lighting conditions are at play. Consequently, the rendering performance of these methods can deteriorate swiftly. For the purpose of this work, we highlight bidirectional methods as they inform the fundamental structure of our design.\\

\textbf{Bidirectional Rendering and Photon Mapping}
In bi-directional path tracing~\cite{veach1995bidirectional}, light paths are not only traced from the camera but simultaneously from the light source. Paths are terminated once a certain threshold of bounces is exceeded or if the ray exits the medium. At this point, the two partial paths ---one from the camera and one from the light source---are connected to form a full path between the viewer and the light source. This guarantees that each path meaningfully contributes to the final rendering. In its pure form, bi-directional path tracing is unbiased.

Photon mapping~\cite{jensen1996global} is a type of biased bi-directional rendering in which the tracing steps are more strictly separated. Paths starting at the light source are traced offline in a pre-computation step. Later, this partial information is used when tracing paths from the camera. The incident radiance at a sampling point can be queried from the previously computed light paths.
In volumetric photon mapping~\cite{jensen1998efficient}, light propagation is simulated by generating and tracing photons from a light source and recording their path through a participating medium. These photon records are stored in what is called a \textit{photon map}. Such maps are usually represented using a spatial partitioning structure like a kd-tree~\cite{bentley1975kdtree} to allow for faster data access and efficient range queries.
When rendering with photon maps, we traverse the volume, estimating the photon density each step of the way. Using these density estimates, we can reconstruct the radiance for each pixel of the image. The density estimation step introduces bias but also smooths the overall radiance contribution, making the method suitable for scenarios in which mostly low-frequency radiance can be expected and where the bias characteristics of the estimator are of secondary concern. 

There are several ways to estimate local radiance in photon maps that depend on the photon representation and sampling technique. Jensen et al.~\cite{jensen1996global,jensen1998efficient} use a k-nearest-neighbor (KNN) range query around a sample point, which remains the simplest and universally applicable method for photon density estimation. Beyond this, Jarosz et al.~\cite{jarosz2008beam} explore beam-shaped queries, replacing point samples. They generalize this approach to include beam-like photons in a follow-up work~\cite{jarosz2011comprehensive}. Finally, Bitterli et al.~\cite{bitterli2017beyond} extend this thought to higher-dimensional space where photons and radiance queries can be represented as points, beams, or even high-dimensional shapes. Generally, higher dimensions yield better results at the same cost. We refer the interested reader to Křivánek et al.'s~\cite{kvrivanek2014unifying} excellent survey of advanced photon density estimation techniques. A common caveat of these higher-dimensional estimators is that they require knowledge of the transmittance properties of a medium. On the one hand, this makes them ideal for homogeneous media as the transmittance can be determined analytically. In heterogeneous media, however, we have to resort to ray marching or monte carlo integration to determine transmittance, which diminishes the returns of such methods. In our work, and as a proof of concept, we choose KNN queries due to their simplicity and universality when it comes to non-homogeneous media. Futhermore, we discuss potential ways to incorporate higher-dimensional estimates in future work.\\

\textbf{Neural Rendering and Neural Fields}
The rapid development of artificial intelligence research has given rise to many exciting applications in the graphics and visualization domain. Naturally, we only cover a relevant fraction of this body of work here.

In the field of scientific visualization, there are several notable examples of how we can address crucial challenges through the use of machine learning. The works by Weiss et al.~\cite{weiss2021neuralvoliso, weiss2022neuraladaptive} introduce screen-space methods to learn super-resolution, sampling, and reconstruction of different volume visualization methods. Bauer et al.~\cite{bauer2023fovolnet} introduce a method---also operating in screen space ---that enables faster volume rendering through sparse sampling and neural reconstruction. Finally, Engel et al.~\cite{engel2020deep} introduce a network that learns a transfer-function-aware neural representation of a dataset's ambient occlusion volume, allowing faster rendering through single-shot inference. Lu et al.~\cite{lu2021compressive} and many others~\cite{han2022coordnet, wu2022instant, weiss2021fast} developed volumetric neural representation that can reduce the size of volume data by over $1000\times$ while still preserving high-frequency details. We refer to Wang et al.~\cite{wang2022dl4scivis} for a comprehensive survey in this area.

Aside from data visualization, neural rendering---as it is termed now---has found applications in almost every area of computer graphics. We highlight several works related to this project, specifically those targeting indirect global illumination. For path tracing, Kallweit et al.~\cite{kallweit2017deep} develop a neural network for offline path tracing that can predict incident radiance in cloud-like volumes. It works by building local descriptors at volume sampling points which are then processed by a neural network. For scenarios without participating media, Gao et al.~\cite{gao2022neural} propose a screen-space method to predict the indirect illumination in a scene based on direct illumination and auxiliary features like scene depth. Lastly, Zhu et al.~\cite{zhu2020deep} address issues in photon mapping. They build a neural radiance estimator that replaces conventional photon density estimates discussed in the previous section. Through this they achieve similar quality with significantly fewer traced photons.

\begin{figure*}[!t]
    \centering
    \includegraphics[width=0.85\textwidth]{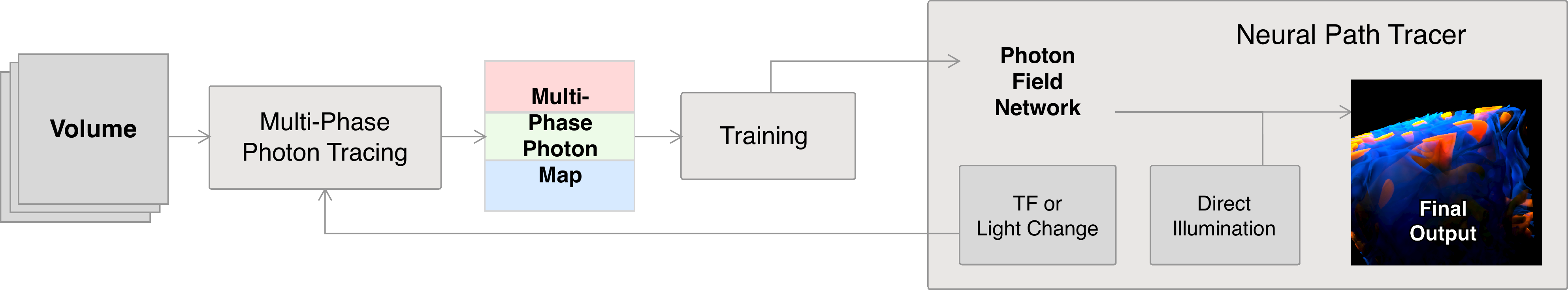}
    \caption{Overview of our pipeline. First, multi-phase photon maps are traced and used to train the photon field network. The field is used during path tracing to provide phase-function-dependent indirect illumination, and it can be retrained on demand when transfer function or lighting conditions change.}
    \label{fig:pipeline}
\end{figure*}

Parallel to these developments, we observe a return to more traditional forms of machine learning models like the multi-layer perceptron. This trend was fueled by the introduction of neural radiance fields~\cite{mildenhall2021nerf}, or NeRFs for short.
At their core, NeRFs are compact multi-layer perceptrons (MLPs) that can implicitly represent dense field data in 5-dimensional function space. In practice, this means that they approximate a function that maps position and view direction to color and density:
\begin{equation}
    F_{\Phi}(x,y,z,\theta,\phi) = (R,G,B,\sigma)
\end{equation}
One of the factors that stalled the successful appropriation of NeRFs in graphics and visualization was the model's practical performance. It entailed long training times and slow inference and rendering speeds. This drawback was soon addressed by Müller et al.~\cite{mueller2021neurradcaching, tiny-cuda-nn}, who introduced a low-level GPU implementation of MLP networks combined with trainable hashgrid encodings~\cite{muller2022instant}. These instant neural graphics primitives~\cite{muller2022instant} were shown to converge within seconds and can be inferred in real-time. Wu et al.~\cite{wu2022instant} introduced such primitives to the field of scientific visualization for interactive visualization of volumetric neural representation.

In this work, we leverage the simple design of NeRFs~\cite{mildenhall2021nerf} and the performance of fast neural representations~\cite{tiny-cuda-nn, wu2022instant} to implement photon fields. Furthermore, we take inspiration from neural rendering approaches~\cite{kallweit2017deep, gao2022neural} that focus on indirect global illumination to speed up the otherwise costly path-tracing procedure.




    \section{Methodology}

Our goal is to provide a way to eliminate costly, data-dependent sampling steps involved in realizing global illumination in scientific volume rendering. We replace this step with a constant-time neural network inference that scales at a much slower pace compared to baseline methods. To this end, we introduce photon field networks that implicitly represent a 6-dimensional illumination volume that can be characterized as a functional mapping from sample position, direction, and phase function coefficient to directional post-integrated photon density. The fields can be used during rendering to substitute the costly evaluation of long scattering paths. Figure~\ref{fig:pipeline} gives an overview of the complete pipeline.

\subsection{Photon Fields}
We introduce the concept of photon fields. These fields approximate the continuous function of local post-estimation radiance density parameterized for the phase function coefficient.

A photon field $P_{\phi}$ establishes the following mapping:
\begin{equation}
    P_{\phi}(x,g,\vec{\omega}) = (R,G,B)
\end{equation}
\noindent where $x \in [0,1]^3$ is a position in space, $g \in [-1,1]$ is a coefficient determining the scattering behavior of light ranging from strong backward ($g=-1.0$) to forward ($g=1.0$) scattering (Figure~\ref{fig:hg_plot}), and $\vec{\omega}$ is a view direction represented in normalized spherical coordinates. 

We propose to use this mapping to represent phase-function-sensitive, indirect, in-scattered radiance (i.e., the indirect global illumination component of the volumetric rendering equation), such that the photon field represents the multiple-scattering term $L_i$ of the rendering equation:
\begin{equation}
    L(x,\vec{\omega}) = L_d(x,\vec{\omega}) + L_i(x,\vec{\omega})
\end{equation}

\noindent where $L$ is the total radiance at point $x$ in direction $\vec{\omega}$, and $L_d$ represents the direct illumination component while $L_i$ stands for the indirect radiance.
Thus, the photon field $P_{\phi}$ can be written as:
\begin{equation}
    P_{\phi}(x,g,\vec{\omega}) \approx L_i(x,\vec{\omega}) = \int_{\Omega_{4\pi}}p_g(x,\vec{\omega},\vec{\omega_i})L(x,\vec{\omega_i})d\omega_i
\end{equation}
Note that the field's parameter $g$ implicitly represents the chosen phase $p_g$. This means the field does not only represent the indirect global illumination but does so for all possible phase-dependent illumination states of a volume on a continuous scale of phase function coefficients.

\subsection{Multi-Phase Photon Tracing}

In order to generate photon fields, we need baseline data. Our approach uses photon maps to represent indirect global illumination at various phase function settings to train our photon fields. The rationale behind choosing photon maps for this work is twofold. One, unlike monte carlo methods, photon mapping, although biased, suffers from little to no stochastic noise. This makes it ideal for our use case since it produces high-quality, noise-free images faster. At the same time, since indirect illumination tends to be lower frequency than direct lighting, the method's bias---caused by the blurring of the radiance estimate---only has a negligible impact on the final image quality. The second reason for choosing photon maps is that they capture the global state of irradiance in a volume, which is by nature view-independent. This quality of photon maps allows us to train a representation just once and use it to render the volume from any viewpoint using arbitrary camera and light scattering parameters. This global state contrasts many methods like path guiding, which are heavily view-dependent and in which the model must continuously adapt to new views on the data.

\begin{figure}[!htb]
    \centering
    \includegraphics[width=1.0\columnwidth]{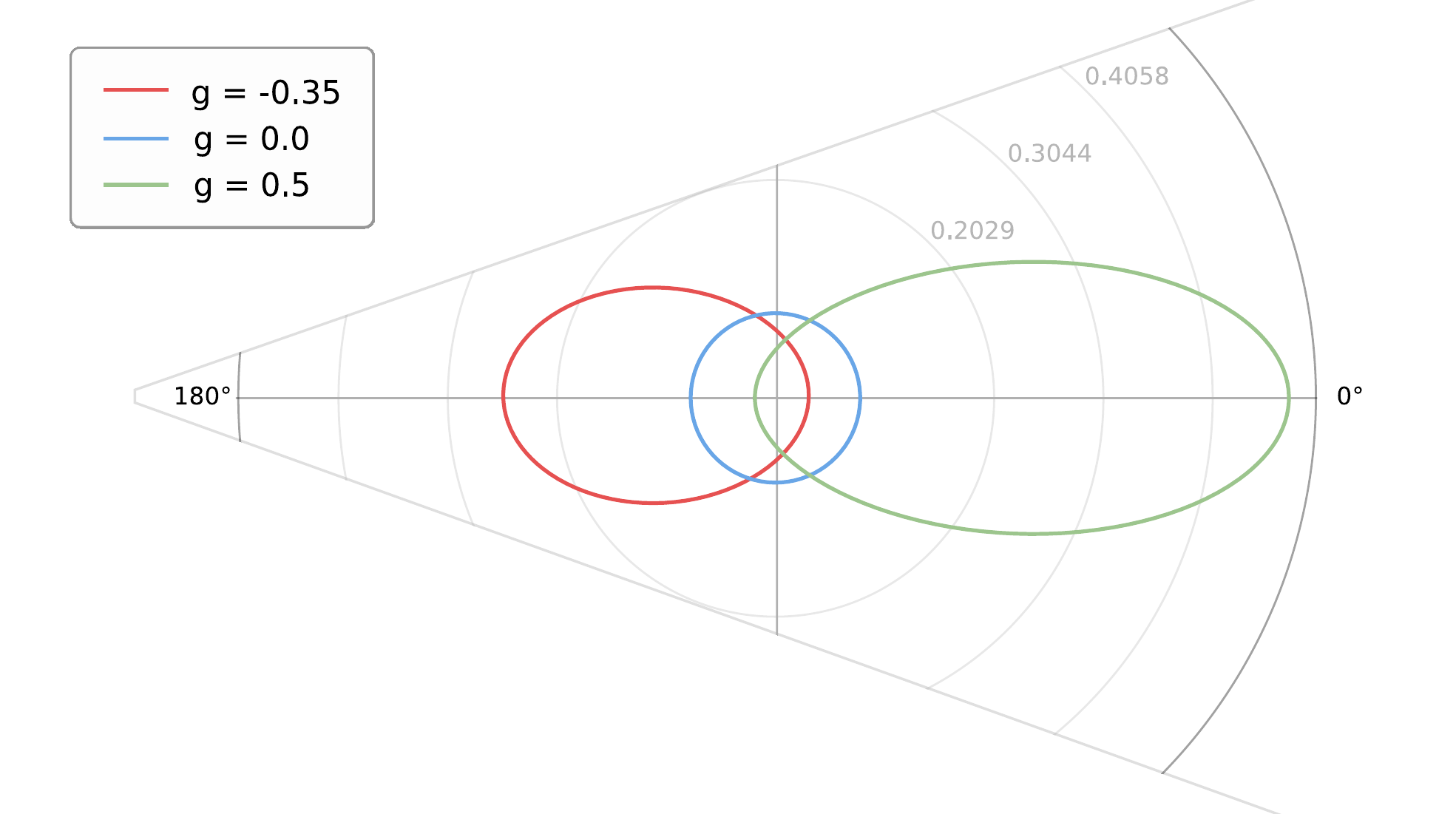}
    \caption{The Henyey-Greenstein phase function~\cite{henyeygreenstein1941phase} at different scattering coefficients for backscattering (red), isotropic scattering (blue), and forward scattering (green). Degrees denote deviation from incoming directions while the radial axis shows the probability density. For instance, at $g=-1.0$ rays have a $p=100\%$ probability of being perfectly reflected while $g=0.0$ means that they will scatter in any random direction ($p=\frac{1}{4\pi}$).}
    \label{fig:hg_plot}
\end{figure}

\begin{figure}[!htb]
    \centering
    \includegraphics[width=1\columnwidth]{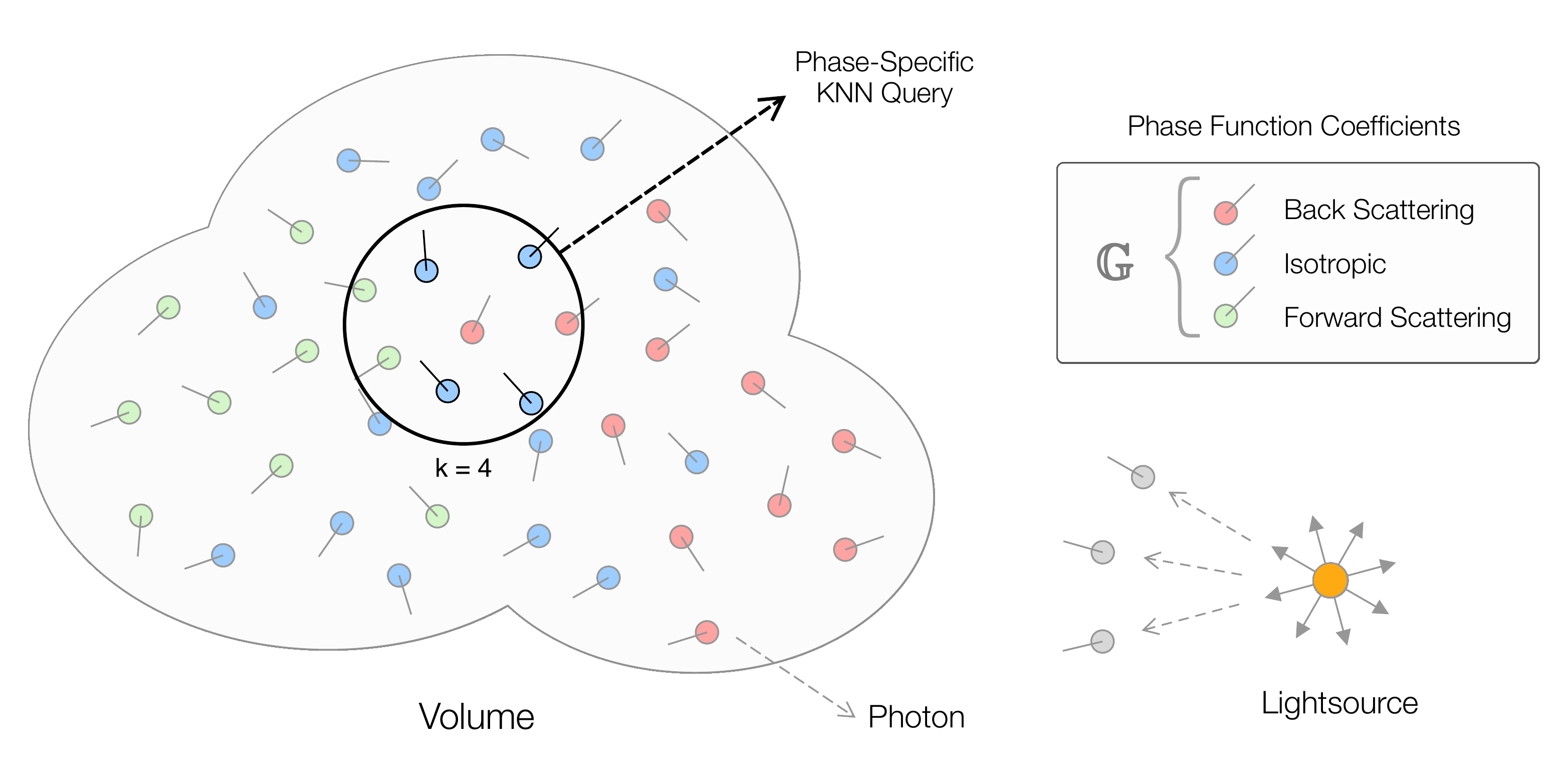}
    \caption{Photons are traced through the volume at discrete phase function coefficients of the Henyey-Greenstein phase function~\cite{henyeygreenstein1941phase}. Photon queries can be performed to filter for a specific type of scattering selectively.}
    \label{fig:photon_tracing}
\end{figure}

To enable the training of phase-variable photon fields, we modify the traditional photon tracing algorithm to work on multiple levels of anisotropy simultaneously. We use the Henyey-Greenstein (HG) phase function~\cite{henyeygreenstein1941phase} since it is easy to compute and allows parameterization of the scattering direction by a single coefficient $g \in [-1,1]$ (Figure~\ref{fig:hg_plot}). The idea is to capture a single photon map that represents the characteristics of multiple phases. To that end, we randomly assign each photon a scattering coefficient $g \in \mathbb{G}$ from a pool of discrete possible coefficients (Figure~\ref{fig:photon_tracing}). The values in $\mathbb{G}$ divide the full phase spectrum into evenly spaced bins. The choice of $\mathbb{G}$ can be made arbitrarily, but we found the set $\mathbb{G} = \{-0.75, 0.0, 0.75\}$ to be the best compromise between tracing and training speed and final field quality.

With these considerations, we generate photon maps by tracing photons from each light source as described in Algorithm~\ref{alg:photontracing}, which illustrates our version of volumetric photon mapping in heterogeneous media. Since the goal is to capture indirect light transport, we only store photons past their first interaction with the medium.

\begin{algorithm}
\caption{Multi-phase photon tracing in heterogeneous media}
\label{alg:photontracing}
\begin{algorithmic}
\State n\_photons $\gets 0$
\State photonmap $\gets \emptyset$
\While{n\_photons $<$ max\_n\_photons}

\State bounces $\gets 0$
\State throughput $\gets 1$
\State photon $\gets $ generate from random light source
\State photon.g $\gets $ random pick out of $\mathbb{G}$
\While{bounces $<$ max\_bounces}
\State compute sample distance via delta tracking~\cite{woodcock1965}
\If{no interaction via delta tracking}
\State \textbf{break} \Comment{photon left the volume}
\EndIf

\State photon.position $\gets$ interaction location
\State photon.direction $\gets $ \textsc{Phase}$(g,$ photon.direction$)$
\State sample $\gets$ \textsc{Volume}$(x,y,z)$
\State albedo $\gets$ \textsc{TransferFunction}$($sample$)$
\State throughput $\gets$ throughput $\cdot$ albedo.alpha $\cdot$ albedo.rgb
\State photon.intensity $\gets$ light.intensity $\cdot$ throughput
\If{bounces++ $\geq 1$} \Comment{skip first interaction}
\State photonmap[n\_photons++] $\gets$ photon
\EndIf
\If{russian roulette termination}
\State \textbf{break}
\EndIf
\EndWhile
\EndWhile
\end{algorithmic}
\end{algorithm}


\subsection{Data Sampling}
Training the photon field networks requires generating appropriate samples that cover the problem input domain $(x,g,\vec{\omega})$. To this end, the previously generated multi-phase maps are transferred to the GPU, where a kd-tree is constructed. We then generate batches of samples. Samples can be generated in parallel, making this step fast enough to facilitate quasi-online training. We use a batch size of $2^{16}$. Each batch contains a uniformly sampled set of grid coordinates $x \in [0,1]$, random uniform directions $\vec{\omega}$, and phase function coefficients $g$ randomly chosen from $\mathbb{G}$. 

\subsubsection{Radiance Estimation}
Using the generated input samples, we compute the ground truth radiance for each sample point. For this, we develop a phase-specific k-nearest-neighbor (KNN) range query around the sample point (Figure~\ref{fig:photon_tracing}). This will only consider photons of phase $g$ and is otherwise equivalent to queries suggested by Jensen et al.~\cite{jensen1998efficient}. The value for $K$ is $1024$ and a maximum query radius $r$ is chosen to limit the number of queried photons as needed. This value is provided by our custom training schedule that will be discussed in Section~\ref{sec:training}.

The final radiance value $L_i(x,\vec{\omega})$ is computed using the KNN adaptive radius radiance estimator~\cite{jensen1998efficient}:
\begin{equation}
    L_i(x,\vec{\omega}) \approx \sum_{n=1}^{K}p_g(x,\vec{\omega},\vec{\omega_n})\frac{\Phi_n}{\frac{4}{3}\pi r^3}
\label{eq:radiance_estimation}
\end{equation}
where $r$ is the distance to the farthest photon found in the KNN query, $\Phi_n \in \mathbb{R}^3$ and $\vec{\omega_n}$ are the $i$-th photon's intensity and direction, and $p_g$ is the HG phase function as determined by the value chosen for $g$.

\subsubsection{Target Encoding}
Neural network training benefits from data normalization. This usually means mapping the network inputs and targets to common intervals like $[-1,1]$ or $[0,1]$. In our case, the inputs are already in this range. However, there are no practical bounds for values of $L_i(x,\vec{\omega})$ (Equation~\ref{eq:radiance_estimation}), and its value is largely dependent on the local volume density and intensity of photons at a given sample point.

Given the number of factors that contribute to the final value of a sample radiance estimate, it is hard to determine the exact global value bounds without densely sampling across the entire input domain. In most cases, this is not feasible due to the relatively high cost of wide-range KNN queries.

To address this issue, we propose a logarithmic network target and output encoding that has the network predict normalized radiance exponents instead of $L_i(x,\vec{\omega})$ directly. These exponents can be easily bounded by determining the maximum desirable floating point precision. They are easy to compute, reversible, and do not overly skew the input domain given reasonable precision constraints.
Assuming that all values for $L_i(x,\vec{\omega}) \geq 0$, we can compute the transformed values $L^{\prime}_i(x,\vec{\omega})$ as follows. 

\begin{equation}
    L^{\prime}_i(x,\vec{\omega}) = 
    \begin{cases}
    -\frac{log_{10}(L_i(x,\vec{\omega}))}{\psi}& \text{if } L_i(x,\vec{\omega}) > \frac{1}{10^{\psi}}\\
    1& \text{otherwise}
    \end{cases}
\end{equation}
where $\psi$ is the maximum desired floating point precision in the number of digits beyond zero. This guarantees that all target values $L^{\prime}_i(x,\vec{\omega}) \in [0,1]$ with a more evenly spaced distance between very bright and dark areas. Values below the precision threshold $\psi$ are implicitly clamped to zero. We found values of $4$ to $6$ for $\psi$ to be sufficient. To restore $L_i(x,\vec{\omega})$, we apply the following inverse transformation:

\begin{equation}
    L_i(x,\vec{\omega}) = 10^{-L^{\prime}_i(x,\vec{\omega}) \cdot \psi}
\label{eq:expdecode}
\end{equation}

The complete sample generation process leaves us with $2^{16}$ of such $[(x,g,\vec{\omega}), L^{\prime}_i(x,\vec{\omega})]$ training tuples. We pass the full batch to the network during training and regenerate a new batch of $2^{16}$ values for every training step.

\subsection{Model}
The underlying machine learning model that represents the photon field comprises several modules. Figure~\ref{fig:network_architecture} shows an overview of the full architecture. There are encoding, network, and decoding components. The encoding layer (Figure~\ref{fig:network_architecture} (left)) decomposes the field's input into positional, directional, and phase components. Positions and directions are transformed using a 3- and 2-dimensional hashgrid encoding~\cite{muller2022instant} respectively. There are other forms of positional encodings like the frequency encoding used by Mildenhall et al.~\cite{mildenhall2021nerf} or the OneBlob encoding suggested by M\"uller et al.~\cite{muller2019neuralimportance}. However, the hashgrid method has been shown to converge faster~\cite{muller2022instant}, and its computational overhead is well within our performance budget. The phase is left untouched as it represents a unitless scale and would not benefit from further encoding. Our encoding uses $16$ grid levels with $8$ features each, a $2\times$ scaling factor between levels, and a hash table size of $2^{19}$. We encode positions and directions separately to decouple their effects on the value of $L_i(x,\vec{\omega})$, which is important for heavily view-dependent phase values $g$.

The network layer (Figure~\ref{fig:network_architecture} (middle)) is a 5-layer perceptron with 64 neurons per layer, ReLU activations between the hidden layers, and no output activation. It takes the encoded inputs from the previous step and processes them to produce a prediction of the log-encoded radiance $L_i^{\prime}(x,\vec{\omega})$. Finally, the exponential decode step (Figure~\ref{fig:network_architecture} (right)) transforms $L_i^{\prime}(x,\vec{\omega})$ to $L_i(x,\vec{\omega})$ using Equation~\ref{eq:expdecode}.


\begin{figure*}[!t]
    \centering
    \includegraphics[width=0.85\textwidth]{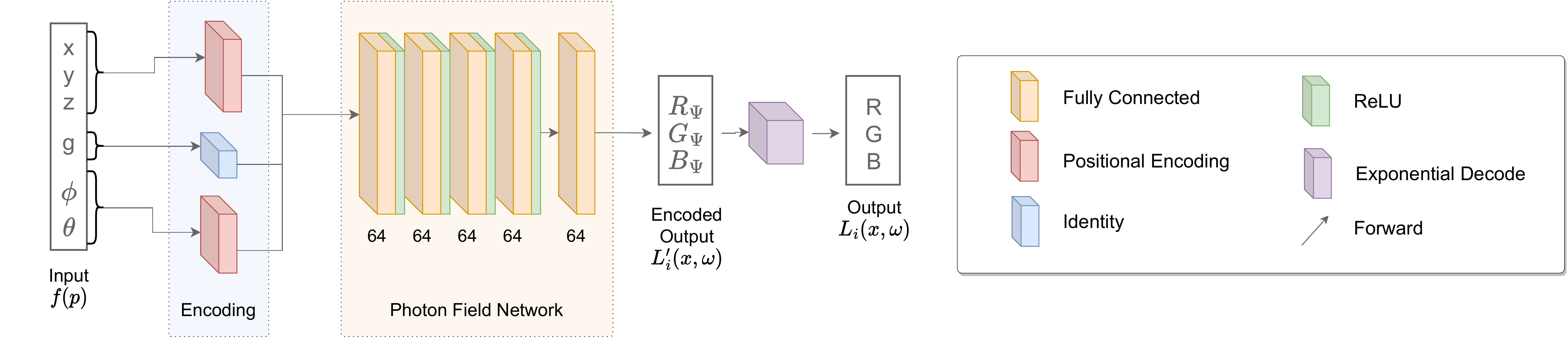}
    \caption{Network architecture of our neural photon fields. 6-dimensional inputs are first separately encoded before being passed to the network. We use the hashgrid method~\cite{muller2022instant} for both positional encodings. The main network is a lightweight MLP with ReLU activation, which produces the log-encoded target values. Decoding them produces the final radiance $L_i$ at $x$ in direction $\vec{\omega}$ at phase coefficient $g$.}
    \label{fig:network_architecture}
\end{figure*}

In contrast to most neural field methods, photon fields do not solely contribute to the appearance of the final image as they produce only the indirect in-scattering term of the rendering equation. These terms tend to be lower-frequency fields and allow for the use of smaller, faster networks compared to approaches that require highly-detailed spatial features.

\subsection{Training}
\label{sec:training}
When it comes to optimizing the photon field, we employ a relatively simple training framework consisting of a single loss term with appropriate hyperparameter settings to regularize training. We develop a custom staggered training schedule to improve the overall time to convergence.

\subsubsection{Hyperparameters and Loss}
During training, we use the Adam~\cite{adam} optimizer with settings for $\beta_1=0.9$, $\beta_2=0.99$, no $L_2$ regularization. The learning rate is set to $9^{-4}$ with an exponential decay of $0.92$ starting at $70\%$ of the total number of training steps and decay triggers at every subsequent $25$ steps. The relative mean square error (rMSE) serves as the training loss, which is applied to each set of outputs individually. We found training sessions between $2000$ and $3000$ steps to be most effective.

\subsubsection{Staggered KNN Schedule}
A downside of the KNN estimator is the need to sample for a relatively large number of photons (i.e., large values for $K$) to generate good estimates for $L_i(x,\vec{\omega})$. Performing these range queries is costly and can quickly deteriorate performance. This is one of the reasons why photon mapping did not enjoy widespread popularity in the past.
We rely on these queries only during training, which somewhat lifts the burden of finding performant estimators. There is, however, a remaining concern about making the training step as fast and unobtrusive as possible. To that end, we propose a staggered training schedule to improve time to convergence. 

\begin{figure}[!htb]
    \centering
    \includegraphics[width=1\columnwidth]{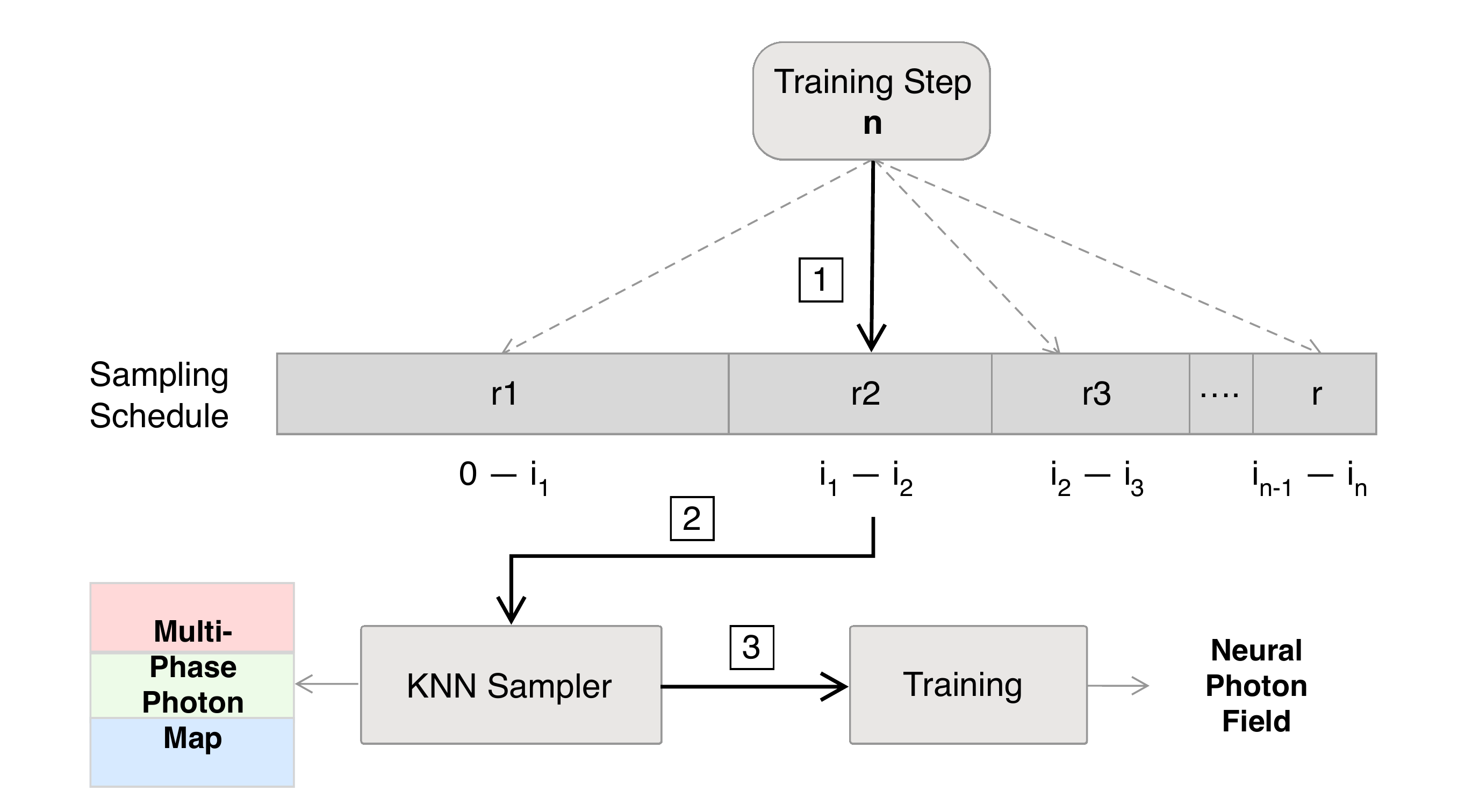}
    \caption{Sampling radii for the KNN queries during training are determined by a custom sampling schedule. The schedule determines the maximum query radius for the current training step (1). The resulting radius (2) is used by the KNN sampler to produce samples (3) which are used for the next training step.}
    \label{fig:sampling_schedule}
\end{figure}

For each photon field, we define a sequence of increasing KNN query radii $r_i$ that approach a target radius $r$. The sequence is queried during training to determine the current step's sampling radius (Figure~\ref{fig:sampling_schedule}). The idea is to quickly accommodate the field to the overall characteristics of the dataset in the early stages with smaller radii and leave the refinement of visual details for later in training. This works because restricting the radius causes the query to return $k < K$ samples given a large enough $K$. The larger the radii become, the more likely it is that $k = K$ at which point the queries are saturated and take the maximum time as constrained by $K$.

\subsection{Rendering}
The trained photon field can be used to render volumes with volumetric global illumination. Since the field represents $L_i$, we can use it to replace the multiple-scattering term in either path tracing or ray marching.

We implement a wavefront path tracer inspired by Wu et al.'s work~\cite{wu2022instant} to render datasets using pre-trained photon fields. The rendering process differs from traditional volume path tracing in that we compute only the first interaction with the volume. We use this set of sample points to estimate direct illumination via delta tracking~\cite{woodcock1965} and query the photon field at the resulting interaction positions to obtain the contribution of indirect global illumination.

Algorithm~\ref{alg:neuralpt} outlines the parallel neural photon field rendering algorithm. The process is designed as a GPU path tracer and is split into multiple device kernels. We provide details about each kernel in the following.

\begin{algorithm}
\caption{Volumetric path tracing with neural photon fields}\label{alg:neuralpt}
\hspace*{\algorithmicindent}\textbf{Input} $camera$, $N$ \\
\hspace*{\algorithmicindent}\textbf{Output} $L$
\begin{algorithmic}
\State $rays \gets \emptyset$
\State $samples \gets \emptyset$
\State $spp \gets N$
\State $rays \gets$ \textsc{RayGeneration} ($camera$)
\State $samples \gets$ \textsc{VolumeSampling} ($rays$, $spp$)
\State $L_i^{pred} \gets $ \textsc{Decode}($P_{\phi}(samples)$)
\State $L_d \gets$ \textsc{DirectIllumination} ($samples$)
\State $L \gets$ \textsc{Compose} ($L_d$, $L_i^{pred}$, $samples$, $spp$)
\end{algorithmic}
\end{algorithm}

\textbf{Ray Generation}
Based on the camera parameters provided to the renderer, we spawn view rays and perform basic intersection testing. If the ray intersects the volume, we store its information in a buffer $rays$, which will be used in later steps.

\textbf{Volume Sampling}
We use the results of the previous kernel to generate $spp$ number of interactions in the volume. Since most scientific volumetric datasets are heterogeneous media, we use delta tracking~\cite{woodcock1965} to estimate the depth of the first interaction.
For each sample on a ray, we use identical ray parameters to simulate multiple "first" interactions with the medium. This is equivalent to performing multiple rendering passes in traditional path tracing and storing only their first interaction.
The maximum number of samples $|samples|_{max}$ generated in this step is $spp \times |rays|$. Samples are stored in a buffer of size $|samples|_{max}$ and invalid interactions are left blank.

\textbf{Photon Field Prediction and Decoding}
The normalized radiance exponent is obtained by inferring the neural field at all values in $samples$. The final predicted radiance $L^{pred}_i$ can be decoded simply by performing the operation described in Equation~\ref{eq:expdecode}. With our design, we are able to batch all $|samples|$ number of samples into a single neural network inference step which significantly reduces computational overhead compared to na\"ive, per-sample or per-ray inference. To improve performance further, we compact the $samples$ buffer, discarding samples where the delta tracking step did not generate a valid interaction, thus removing 'holes' in the uncompacted buffer.

\textbf{Direct Illumination}
Samples generated in the sampling step are also used to sample direct light. We use next-event estimation (NEE) and delta tracking~\cite{woodcock1965} to sample the light sources directly. This is equivalent to performing single-bounce illumination in a traditional path tracer.

\textbf{Composition}
This kernel combines direct and indirect light contributions. It computes the scattering probability $\sigma_s$ at each given sample point in the volume and determines the multiple importance sampling weights (MIS)~\cite{veach1995optimally} for direct lighting and phase function sampling. With these values, it composes the total illumination for each ray as follows.
$$
L = \frac{1}{SPP} \sum_{k=1}^{SPP} w_d L_{d_k} + w_i\sigma_{s_k} L_{i_k}^{pred}
$$
where $w_d$ and $w_i$ are the MIS~\cite{veach1995optimally} weights for NEE and the continued path sampling, which is represented by the network prediction in this case.
This leaves us with the final value for $L$. It can be displayed directly, used for frame accumulation, or denoised to reduce the remaining noise caused by the initial sampling step.
    \section{Evaluation}
To show the merits of photon field networks, we perform several evaluations on different properties of our design.
\subsection{System Setup}
We use C++ backends for both the OpenVKL~\cite{openvkl} and oneTBB~\cite{tbb} libraries to implement the photon tracer. Photon maps are built on the GPU using a version of the cuKD-Tree~\cite{waldcukd} implementation that was adapted by us to support selective queries. We chose this library as it operates stackless and without recursion, allowing for much larger photon maps that exceed the maximum recursion depth or stack size imposed by the GPU. Training and inference of photon fields was realized using the Tiny CUDA Neural Networks~\cite{tiny-cuda-nn} library using fully-fused MLPs at fp16 half-precision. Finally, we implement baseline rendering and sampling methods in both CUDA and OpenVKL~\cite{openvkl}. Neural rendering involving photon field sampling was written in CUDA.
All computations and evaluations, including the photon tracing, training of the photon fields, and rendering the final images was done on an end-user machine with two Intel Xeon Gold 5220 CPUs, 256 gigabytes of RAM and an NVIDIA GeForce RTX 3090. The system runs Ubuntu 20.04 LTS and all parts of the project were compiled using GCC 8.4 through NVCC running CUDA 11.6. All data is recorded at 720p unless otherwise specified.

\subsection{Training Performance}
The initial generation of photon fields is an important part of the proposed pipeline. In this section, we evaluate training speed at different capture settings. Furthermore, we analyze the loss progressions for our staggered training schedule.

\begin{figure}[!htb]
    \centering
    \begin{subfigure}[b]{1\columnwidth}
        \centering
        \includegraphics[width=1\columnwidth]{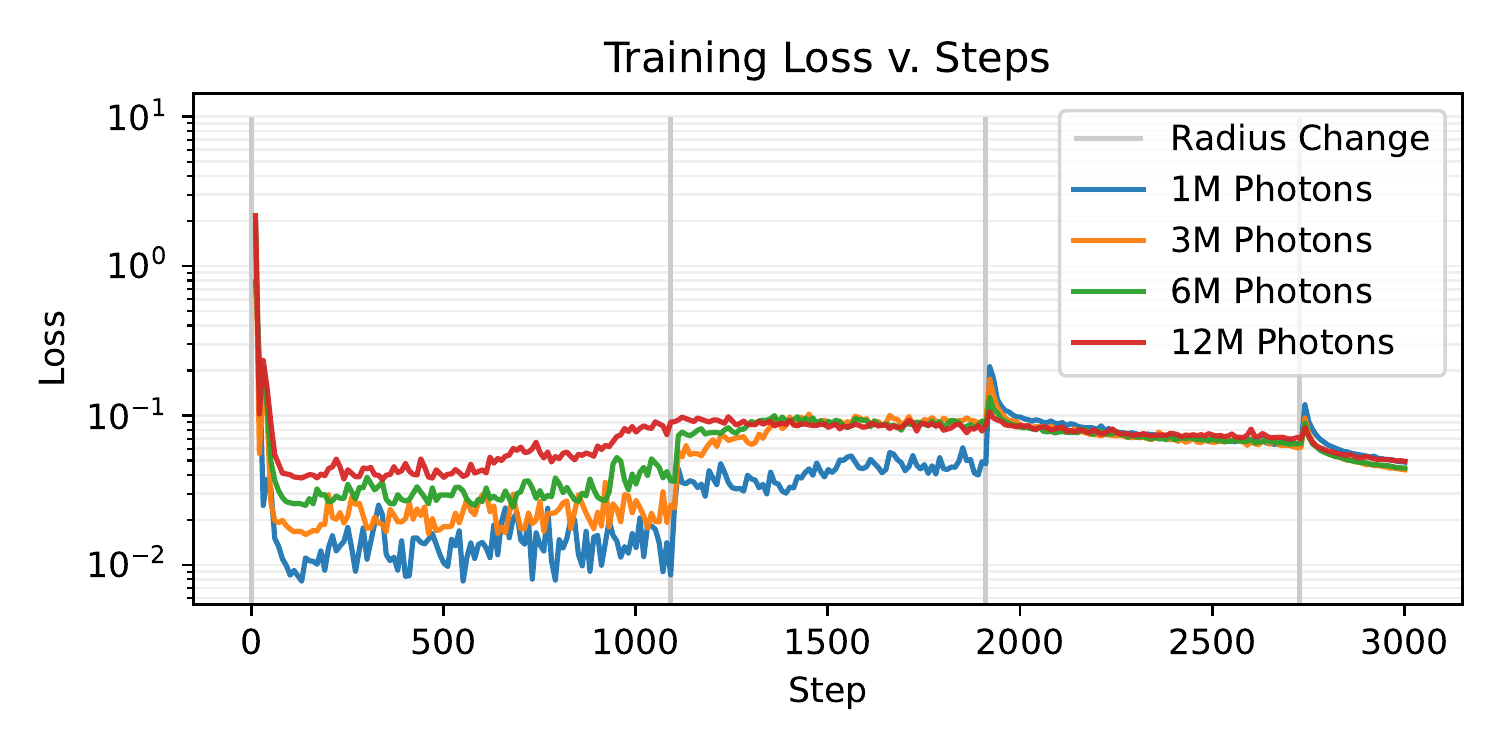}
        \vspace{-3.50em}
     \end{subfigure}
     \hfill
     \begin{subfigure}[b]{1\columnwidth}
        \centering
        \includegraphics[width=1\columnwidth]{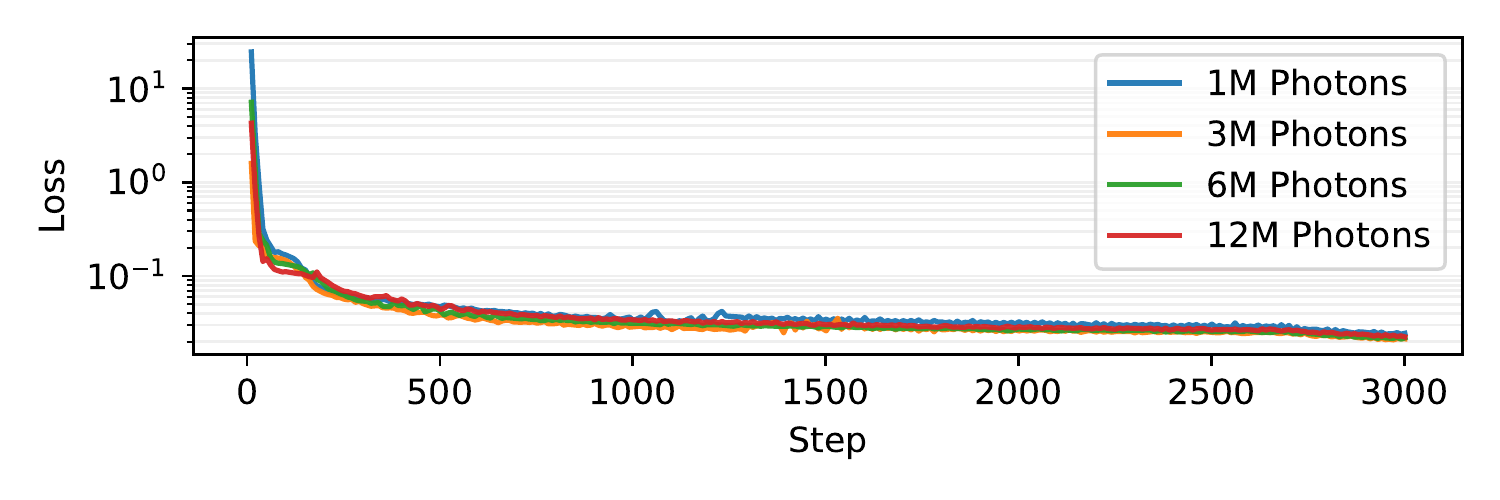}
     \end{subfigure}
    \caption{Example of a representative training on the \textit{Vortices} dataset. The photon field is trained for $3000$ steps at $2^{16}$ samples per step. Training is performed on multiple maps with increasing number of total photons. The top chart shows our staggered training with vertical lines indicating radius changes while the bottom shows a naïve run. Both reach comparable final quality while the staggered approach takes significantly less time to complete.}
    \label{fig:training_steps}
\end{figure}

\begin{figure*}[!htb]
    \centering
    \includegraphics[width=1\textwidth]{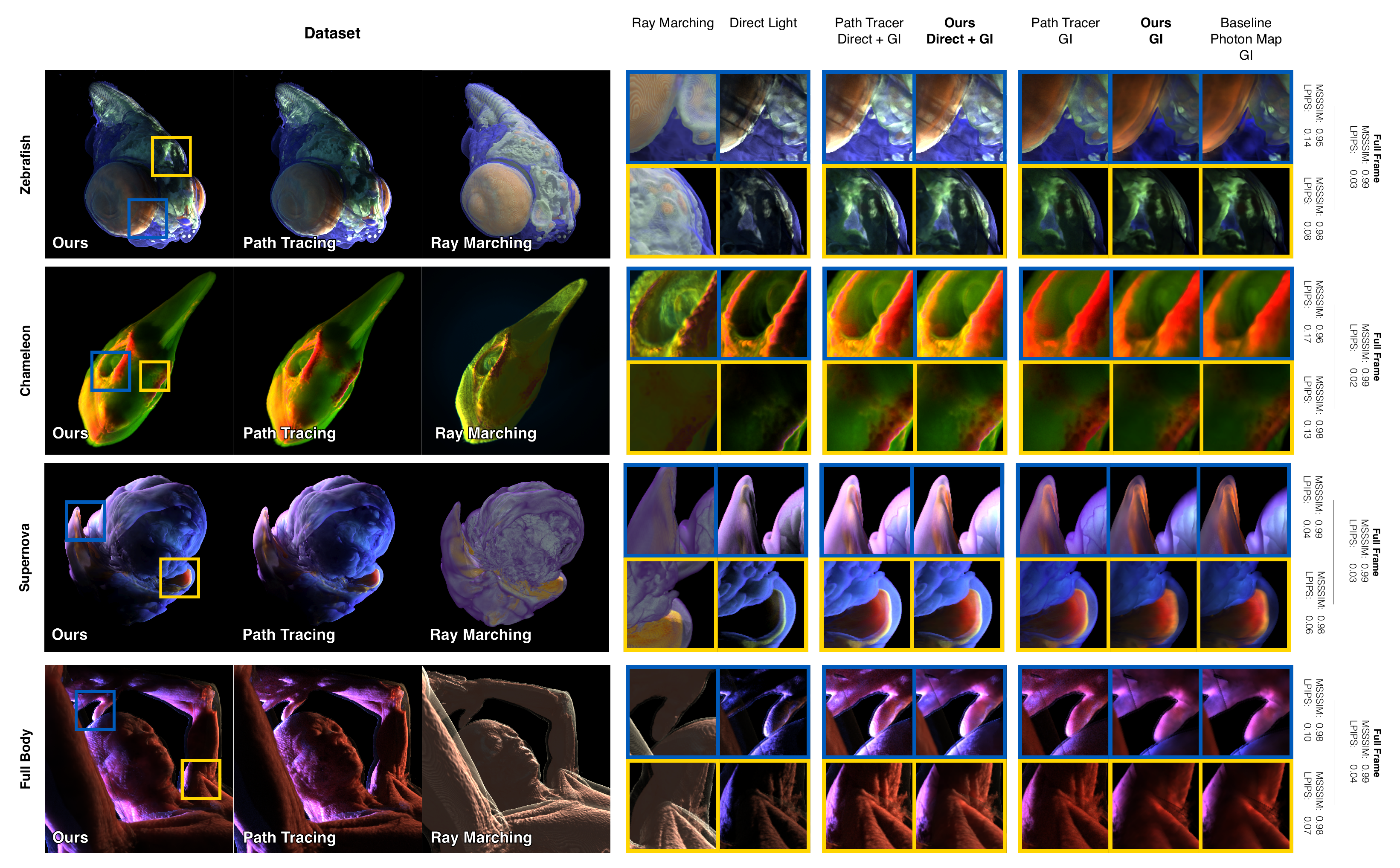}
    \caption{Visual quality of our approach compared to two rendering methods often used in scientific visualization---path tracing and ray marching. We show decomposed views to highlight the contribution of $L_i$ and how it compares to the baseline photon mapping and path tracer. Overall, our method faithfully represents the baseline while successfully emulating the look of the path tracer.}
    \label{fig:quality}
\end{figure*}

First, we show data for a representative training on the \textsc{Vortices} dataset. The training was performed on multiple photon maps containing increasing number of photons. Figure~\ref{fig:training_steps} shows data for trainings on $1M$, $3M$, $6M$, and $12M$ photons. The number of photons denote the total number of particles in the map, which means there are $\frac{x}{|\mathbb{G}|}$ photons per phase. Here, we chose a $40/30/30/10$ split for the staggered training schedule. Schedule shifts are indicated by vertical lines and specific target radii are listed in Table~\ref{tab:training_target_radii}. The radii are adjusted to roughly match the data size; however, we do not prescribe optimal settings for this step. They can be changed as needed to balance quality and training time. Longer trainings warrant the choice of more intermediate radii. Ultimately, we found the exact design of the KNN schedule to be an empirical matter which largely depends on the dataset, lighting conditions, and transfer function. Please find the full list of training data, loss curves, and KNN schedules for all datasets in the supplemental material.

\begin{table}[!htb]
    \caption{Example of a four-step staggered KNN schedule. Values shown here were used to conduct the training in Figure~\ref{fig:training_steps}.}
    \centering
    \begin{tabularx}{\columnwidth}{l|X|X|X|X}
    \toprule
        \textbf{Photons} & 0--36\% & 37--63\% & 64--90\% & 91--100\% \\
    \midrule
    \midrule
        \textsc{1M}     & $0.25$ & $0.50$ & $2.50$  & $5.0$ \\
        \textsc{3M}     & $0.25$ & $0.50$ & $2.00$  & $4.0$ \\
        \textsc{6M}     & $0.25$ & $0.50$ & $1.50$  & $3.0$ \\
        \textsc{12M}    & $0.25$ & $0.50$ & $1.00$  & $2.0$ \\
    \toprule
    \end{tabularx}
    \label{tab:training_target_radii}
    \vspace{-0.1in}
\end{table}

The bottom of Figure~\ref{fig:training_steps} includes a comparison to a naïve training run. At first sight, both trainings seem to arrive at similar losses and produce similar visual quality. However, timing data shown in Figure~\ref{fig:training_time} reveals that the staggered training takes only a fraction of the time it takes the naïve approach to complete. Over all datasets, we see approximately $2$--$3\times$ improvement in training speed using our staggered training.

\begin{figure}[!htb]
    \centering
    \includegraphics[width=0.85\columnwidth]{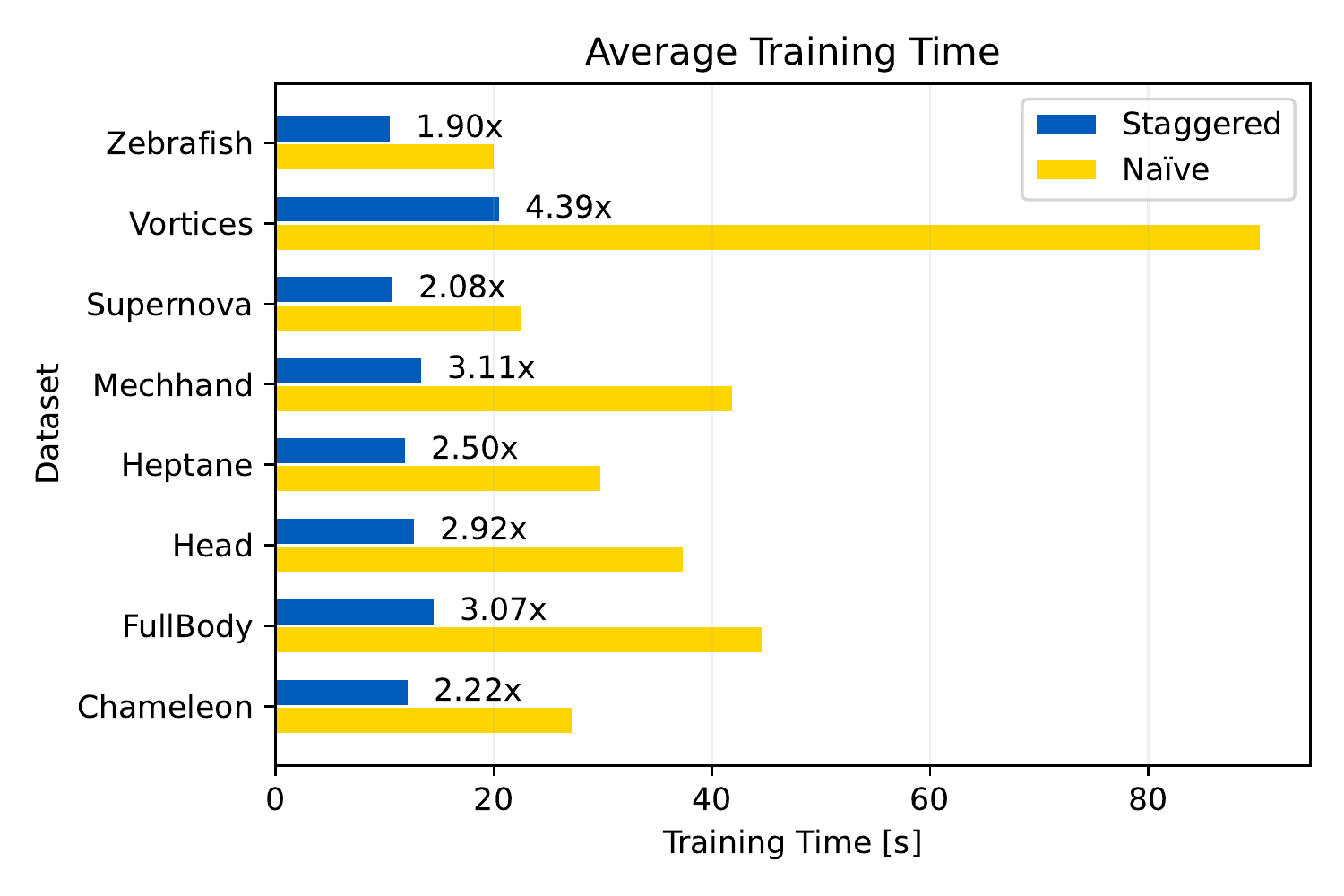}
    \caption{Average timings of trainings on all datasets over 3000 steps comparing the effects of using our KNN schedule on training time. Naïve trainings (yellow) were conducted using the target radius $r$ directly, while our schedule (blue) approaches $r$ incrementally.}
    \label{fig:training_time}
\end{figure}

\subsection{Rendering Results}
In this section we evaluate the performance and visual quality of our method. To put them into context, we compare our results against several baseline methods that are commonly found in scientific visualization. First, we consider a direct comparison with the underlying photon mapping method to showcase the reconstruction quality of our fields. Furthermore, we show data from a ray marcher with single-light global shadows. Lastly, we implement a path tracer with full global illumination using next-event estimation (NEE) and multiple importance sampling (MIS) to show how our method compares to the effects that it aims to emulate.

\subsubsection{Rendering Performance}
To determine the performance of our method we gather the per-frame timings of the rendering pipeline components. Data is recorded at 1SPP across all datasets. The timings are an average over a 2000-frame fly-through sequence that captures the dataset from various angles and at different distances from the volume. We measure the relative speedup of generating $L_i$ and $L$ using our photon field method as compared to a full path tracer. The data is shown in Table~\ref{tab:rendering_performance}. All values were generated with a maximum path length of $16$. Data for the underlying photon mapping rendering pipeline is omitted as the KNN queries are magnitudes slower than both the path tracer and our method.

The data shows that by cutting paths short and instead using the photon field to determine $L_i$ saves a significant amount of time. Our approach outperforms the path tracer by $4$--$5\times$ on average and by as much as $13\times$ when considering only the relevant sub-process of generating the indirect illumination term. Throughout most datasets, we observe a strong correlation between the screen fill ratio of the rendered volume to the overall frame time speedup. We attribute this to the exponential difference in the number of secondary rays used to path trace $L_i$ compared to the linear increase in network inferences when using our photon fields. 

\begin{table*}[!htb]
    \caption{Render timings averaged over a screen-filling, 2000-frame fly-through sequence. We compare our photon field renderer against a conventional path tracer. Values for $L$ include additional time for ray generation and other smaller processses.}
    \centering
    \begin{tabularx}{\textwidth}{l|l|X|X|X|X|X|c|c}
    \toprule
        \multirow{2}{*}{\textbf{{Dataset}}} & \multicolumn{3}{c|}{\textbf{Ours}} & \multicolumn{3}{c|}{\textbf{Path Tracer}} & \multirow{2}{*}{\textbf{{Speedup $L$}}} & \multirow{2}{*}{\textbf{{Speedup $L_i$}}}\\
        & \textbf{{$L_d$ $[ms]$}} &	 \textbf{{$L_i$ $[ms]$}} &	 \textbf{{$L$ $[ms]$}} &	 \textbf{{$L_d$ $[ms]$}} &	 \textbf{{$L_i$ $[ms]$}} &	 \textbf{{$L$ $[ms]$}} & & \\

    \midrule
    \midrule
        \textsc{Chameleon}  &   14.56 &	 4.57 &	 19.40 &	 16.98 &	 57.56 &	 74.74 &	 3.85$\times$ & 12.59$\times$ \\
        \textsc{FullBody}   &   9.18    &	 5.03 &	 14.45 &	 8.95 &	 66.33 &	 75.47 &	 5.22$\times$ & 13.18$\times$ \\
        \textsc{Heptane}    &   6.95    &	 4.92 &	 12.16 &	 6.84 &	 67.26 &	 74.33 &	 6.11$\times$ & 13.67$\times$ \\
        \textsc{Mech.Hand}  &	3.73  &	 5.35 &	 9.40 &	 3.69 &	 35.72 &	 39.63 &	 4.22$\times$ & 6.68$\times$ \\
        \textsc{Supernova}  &	5.69  &	 5.57 &	 11.55 &	 5.61 &	 58.87 &	 64.68 &	 5.60$\times$ & 10.57$\times$ \\
        \textsc{Vortices}   &	1.56  &	 5.74 &	 7.57 &	 1.85 &	 14.47 &	 16.61 &	 2.19$\times$ & 2.52$\times$ \\
        \textsc{Zebrafish}  &	7.87  &	 4.29 &	 12.45 &	 7.77 &	 52.37 &	 60.32 &	 4.85$\times$ & 12.22$\times$ \\

    \midrule
        \textbf{\textsc{Overall}} &	         \textbf{7.08} &	         \textbf{5.07} &	         \textbf{12.42} &	         \textbf{7.38} &	         \textbf{50.37} &	         \textbf{57.97} &	         \textbf{4.58$\times$} & \textbf{10.20$\times$} \\

    \toprule
    \end{tabularx}
    \label{tab:rendering_performance}
    \vspace{-0.1in}
\end{table*}

\begin{figure*}[!htb]
    \centering
    \includegraphics[width=0.95\textwidth]{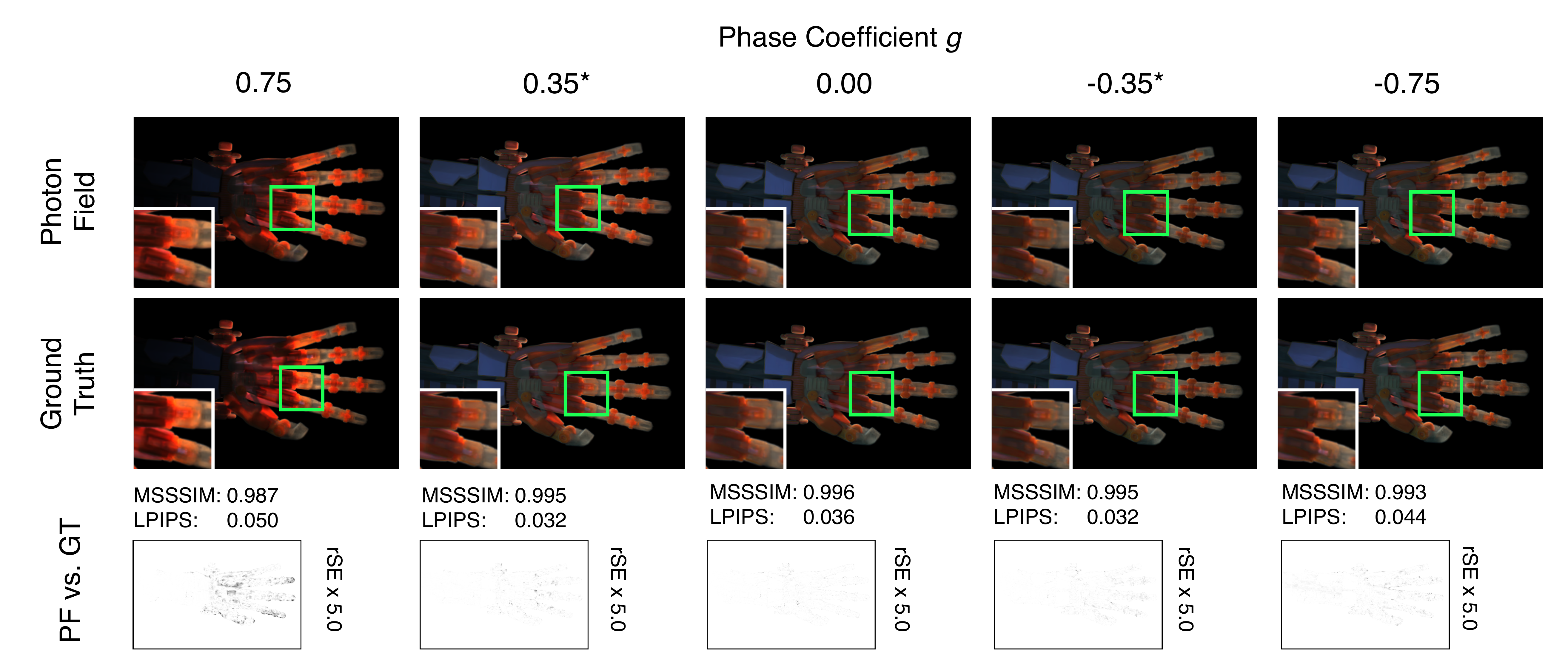}
    \caption{Phase function awareness analysis for photon field networks trained on the \textsc{MechanicalHand} dataset using $\mathbb{G} = \{-0.75, 0.00, 0.75\}$. We show the neural rendering output (first row) and the ground truth data (second row). Data for $g\* \in \{-0.35, 0.35\}$ was not part of the training set $\mathbb{G}$. Each pair of images is evaluated using MSSSIM, LPIPS, and relative square error (third row).}
    \label{fig:phase}
\end{figure*}

\subsubsection{Image Quality}
To evaluate the visual quality of our neural path tracer, we compare it against the baseline photon maps and place it side-by-side with a conventional path tracer and ray marcher (Figure~\ref{fig:quality}). Decomposed views of the data are provided to highlight the contribution of $L_i(x,\vec{\omega})$. We compute perceptual metrics comparing $L_i$ of the baseline photon map against the field's output. This test is indicative of the quality of the network. We abstained from computing metrics comparing the method against the path tracer or ray marcher since they are different rendering methods. While photon mapping---and in extension our photon fields approach---emulates the visual characteristics of path tracing, it is not equivalent. Thus, the metrics do not constitute a fair apples-to-apples comparison. Note that the maximum quality of our method is limited by the quality of the multi-phase photon maps. In this study, we use data from $12$M photon maps, with $4$M photons per phase which provides fairly good quality but comes at a one-time cost when computing the maps (Section~\ref{sec:photon_tracing}). We will discuss how future work can handle this quality constraint more gracefully and with less pre-computation time. 
The quality study results in Figure~\ref{fig:quality} indicate that our approach is able to faithfully reconstruct the radiance represented by the baseline, and despite learning the bias inherent to photon mapping, it closely matches the results of the full volumetric path tracer. For a more exhaustive collection of visual analyses, please refer to the supplemental material.

\subsubsection{Stochastic Noise}
Another advantage of our approach is that it does not suffer from stochastic noise beyond the initial delta tracking sampling step. This results in a much better noise profile when compared to the conventional path tracer. In Figure~\ref{fig:noise} we show a crop from the \textsc{Chameleon} dataset captured at different SPP and compare our approach against the path tracer. It becomes clear that at the same number of samples, photon fields suffer from significantly less noise. Note how our method is consistently better than the path tracer even when doubling or quadrupling the SPP (i.e. compare path tracer at 4 and 8 SPP against ours at 1 and 4 SPP). Generally, we observe that the greater the contribution of $L_i$ and the smaller the contribution of $L_d$ to a pixel's $L$, the more significant the benefit of this effect becomes. As SPP increase, both methods approach similar levels of quality. This fact allows us to produce high quality images much faster, as lower SPP are needed to reach acceptable levels of convergence. Similarly, in cases where a denoiser is employed, our method will produce higher quality sooner.

\begin{figure}[!htb]
    \centering
    \includegraphics[width=1\columnwidth]{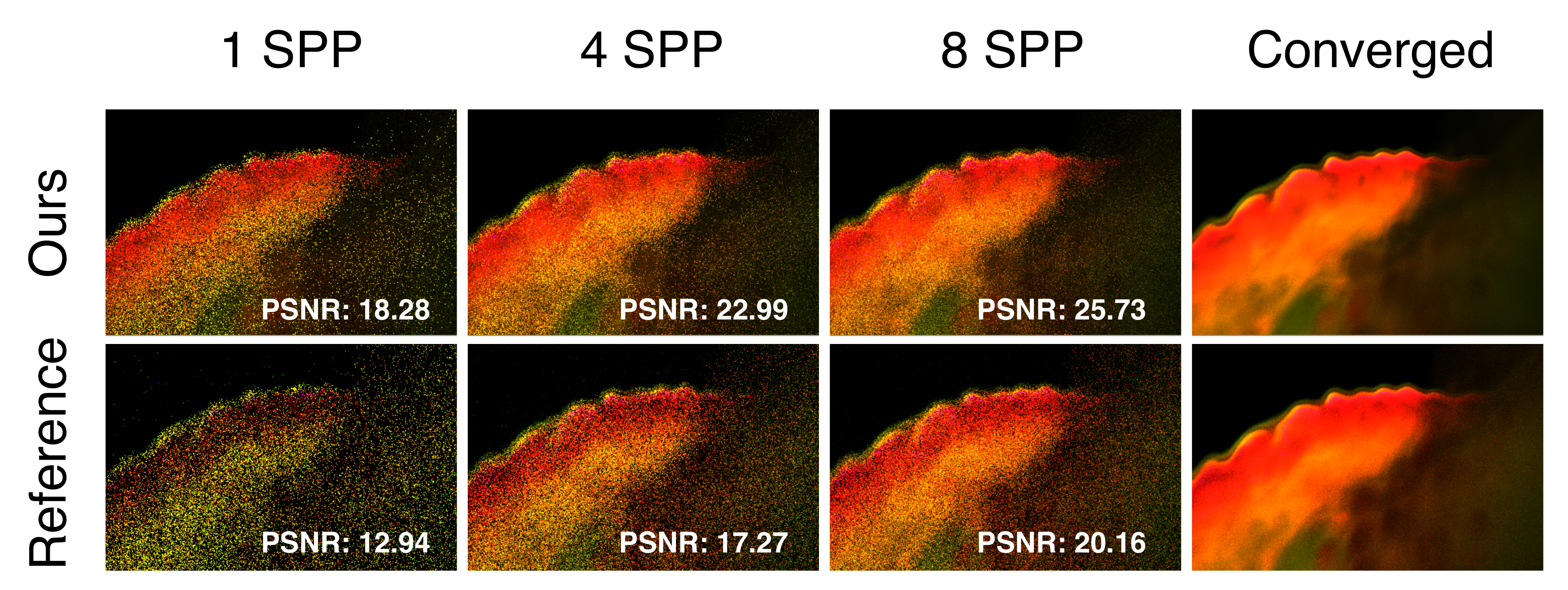}
    \caption{The crest around the \textsc{Chameleon}'s eye cause notable scattering and therefore indirect illumination. The scene is captured at $1$, $4$, $8$, and $1024$ SPP using our method and the reference path tracer. The photon fields exhibit noticeably less noise at lower SPP.}
    \label{fig:noise}
\end{figure}

\subsubsection{Phase Function Awareness}

Photon fields capture $L_i(x,\vec{\omega})$ along the continuum of phase function coefficients $g \in [-1,1]$. To evaluate their capability of capturing the mapping, we compare outputs of a field trained on a set $\mathbb{G}$. We show outputs for $g \in \mathbb{G}$ as well as $g\* \notin \mathbb{G}$. Results for this test are shown in Figure~\ref{fig:phase}. The starred values indicate $g \notin \mathbb{G}$, highlighting the photon field's capability to generalize across the phase spectrum as shown by the structural and perceptual image metrics. Note that there is an inherent error when comparing to unseen data as the photon field learns the bias of the initial set $\mathbb{G}$ (Figure~\ref{fig:phase} (rSE)). Due to the stochastic and biased nature of photon tracing, we find this error even when comparing equivalent baseline images derived from different traces. This means that the results for $g \notin \mathbb{G}$ are a compound of prediction and photon tracing errors. Fortunately, the latter does not drastically impact the perceptual quality of the resulting images as indicated by the other metrics (MSSSIM and LPIPS). Please refer to the supplemental material for an evaluation among equivalent baseline photon maps to showcase the tracing error. We discuss potential future directions to mitigate the influence of estimator bias and tracing error in Section~\ref{sec:discussion}. To get a better sense of how the phase function parameterization influences the visual output, please refer to the supplemental video.

\subsection{Photon Tracing Performance}
\label{sec:photon_tracing}
Creating multi-phase photon maps is an a priori step necessary to build the photon field. Ideally, high-quality photon tracing is employed sparsely, as it constitutes the most cost-intensive task in the whole process. Although not part of our contribution, we deem it important to understand the initial cost of generating baseline data. We do not claim optimality for this step. We discuss potentials for interactive tracing in Section~\ref{sec:discussion}. The deciding factors for tracing speed are the volume's size, density distribution, and the number of photons traced. We have little influence over the former two, so our evaluation focuses on the latter. Table~\ref{tab:tracing_speed} shows an overview of tracing times for different datasets. All times were generated on the CPU using a parallelized tracing procedure. 
The number of photons indicated is the total number of photons traced in the scene. The photons are equally distributed over the number of lights $|\mathbb{L}|$ and the number of phases $|\mathbb{G}|$. We see that, as expected, number of photons strongly correlates with target time. However, the volume size is not always indicative of tracing time due to varying density distributions and the resulting differences in trace lengths at different phase coefficients. To achieve good results, traces of 1M up to 12M photons are necessary to achieve high visual quality for the photon fields. Traces can be created a priori and stored on disk for repeated use during trainings.

\begin{table}[!htb]
    \caption{Tracing speed at different capture settings. We show the tracing speed over all datasets used in the study and report numbers at different levels of quality ranging from 1M to 12M photons.}
    \centering
    \begin{tabularx}{\columnwidth}{l|l|r|r|r|r}
    \toprule
        
        \multirow{2}{*}{\textbf{Dataset}} & \multirow{2}{*}{\textbf{Size}} & \multicolumn{4}{c}{\textbf{Tracing Time [s]}}\\ 
        & & 1M & 3M & 6M & 12M \\
    \midrule
    \midrule
        \textsc{Vortices}   & $128\cdot 128\cdot 128$ & 4.2 & 6.2 & 23.8 & 24.4\\
        \textsc{Heptane}    & $302\cdot 302\cdot 302$ & 4.9 & 13.2 & 29.5 & 54.3\\
        \textsc{Mech.Hand}   & $640\cdot 220\cdot 229$ & 4.1 & 9.2 & 23.9 & 40.1\\
        \textsc{Supernova}  & $432\cdot 432\cdot 432$ & 12.2 & 35.8 & 72.7 & 142\\
        \textsc{ZebraFish}  & $592\cdot 413\cdot 956$ & 23.8 & 70.9 & 140.5 & 282\\
        \textsc{FullBody}   & $512\cdot 512\cdot 1299$ & 4.6 & 13.3 & 26.8 & 52.6\\
        \textsc{Chameleon}  & $1024\cdot 1024\cdot 1080$ & 7.6 & 19.6 & 38.6 & 76.7\\
    \toprule
    \end{tabularx}
    \label{tab:tracing_speed}
    \vspace{-0.1in}
\end{table}



    \section{Limitations and Future Directions}
\label{sec:discussion}
Photon field networks perform exceedingly well in a wide range of scenarios. Naturally, there are limitations to the approach. In this section we identify and discuss some of them. We provide perspectives on how to overcome these limitations and identify potentials to further extend our method.

\textbf{Low Opacity Volumes and Boundary Gradients}
One drawback of photon tracing in conjunction with delta tracking and KNN radiance estimation is its failure to capture illumination in highly transparent regions and areas with high density gradients such a air-surface boundaries. This is due to the limited number of photons we can process and, in turn, the estimators nature of averaging over the entire spherical domain around a sample. This can lead to areas receiving less average radiance than needed to faithfully represent them when there are few or no photons in their vicinity. We identify this as a limitation of the underlying method and, in the following paragraphs, discuss potentials to extend this work by employing more efficient estimation techniques that will at the same time improve performance.

\textbf{Faster Photon Map Retracing}
One bottleneck of our approach is the regeneration of the multi-phase photon traces when scene parameters like transfer function or lighting change. One possible direction to address this issue is to only retrace the parts of the map that have become sufficiently different as a consequence of the scene adjustments. Such an approach has been proposed for traditional photon mapping by Jönsson et al.~\cite{jonsson2016correlated} and we are confident that this can be adapted for our multi-phase maps in the future.

\textbf{Beyond Point Samples}
Another way to reduce computational costs associated with tracing and sampling photon data is to reduce the need for high photon counts. As discussed in related work, some of the higher-dimensional estimates require more work to be adapted for an approach like this. One possible candidate for an extension to this work could be the beam radiance estimate~\cite{jarosz2008beam,jarosz2011comprehensive} which gathers radiance along a ray instead of a sampling point. Subsequently, our photon fields could be adapted to work as a form of light field network~\cite{suhail2022light,sitzmann2021light} which, instead of predicting point samples, is trained to produce per-ray radiance data. At the same time, neural estimators like the one proposed by Zhu et al.~\cite{zhu2020deep} achieve high quality with significantly less photon samples. This can be used during photon field training to manage effectively with much smaller maps. These directions would allow re-tracing and fine-tuning of an existing field at interactive rates.

\textbf{Opportunities Beyond Rendering}
In this paper, we show the merits of photon field networks in volumetric rendering applications. However, there are potential uses for this approach outside of image generation. Most notably, we see considerable need for fast radiance estimates in the field of path guiding. In this line of research, radiance samples are used to optimize scattering distributions for improving the quality of path tracing samples and reducing overall variance of the estimator~\cite{Vorba2019PathG, Herholz2019}. Using photon fields can support efforts that rely on photon traces like Herholz et al.'s work~\cite{Herholz2019} by providing fast estimates that are trained once and can be evaluated in a fraction of the time it would take conventional estimators to do the same. 
    \section{Conclusion}
We introduce photon field networks---a neural rendering method to enable real-time global illumination for scientific volume data. The fields can be trained in seconds, and allow for significantly faster rendering compared to conventional methods like ray marching or path tracing. Our design produces high visual quality with low levels of stochastic noise even at low sample rates. At the same time, the photon fields capture the global illumination along a scale of anisotropic scattering. We show that photon field networks are able to produce global illumination effects as much as $13\times$ faster than a comparable path tracer, all the while producing images that closely match the more expensive approach. In future iterations, we plan to extend our design to accommodate more cost-effective photon estimation techniques like the beam radiance estimate~\cite{jarosz2008beam,jarosz2011comprehensive} to further improve rendering times and reduce the cost of baseline data generation. This will pave the way to near-seamless relighting with all the benefits that already come with photon fields.

In this work, our goal was to demonstrate how machine learning methods can enhance scientific visualization. We see encouraging results from adjacent fields like computer graphics and there are countless opportunities to extend and augment our community's well-established rendering paradigms. We hope to motivate the visualization community to pursue this direction and push the boundaries of what is possible in real-time, high-fidelity volume visualization.

    \bibliographystyle{abbrv-doi}

    \bibliography{template}
\else
    \input{sections/supplemental}
    
\fi

\end{document}